\documentclass[english,aip,reprint,super,sort,compress,onecolumn,longbibliography]{revtex4-1}

\usepackage{ORCIDinREVTeX}
\usepackage{amsmath}
\usepackage{graphicx}
\usepackage{hyperref}
\usepackage{cleveref}
\usepackage{threeparttable}
\usepackage{subfig}
\usepackage{chemfig}
\usepackage{placeins}
\usepackage{comment}

\makeatletter

\@ifundefined{textcolor}{}
{%
 \definecolor{BLACK}{gray}{0}
 \definecolor{WHITE}{gray}{1}
 \definecolor{RED}{rgb}{1,0,0}
 \definecolor{GREEN}{rgb}{0,1,0}
 \definecolor{BLUE}{rgb}{0,0,1}
 \definecolor{CYAN}{cmyk}{1,0,0,0}
 \definecolor{MAGENTA}{cmyk}{0,1,0,0}
 \definecolor{YELLOW}{cmyk}{0,0,1,0}
}

\usepackage{chemfig}
\usepackage[version=3]{mhchem}
\usepackage{braket}
\usepackage{placeins}

\makeatother

\usepackage{babel}
\begin{document}

\newcommand*\citeit[1]{\citeauthor{#1}\cite{#1}}
\newcommand*\citeref[1]{ref. \citenum{#1}}
\newcommand*\citerefs[1]{refs. \citenum{#1}}
\newcommand*\ie{\emph{i.e.}}
\newcommand*\vs{\emph{vs.}}
\newcommand*\etc{\emph{etc.}}

\newcommand*\SL[1]{{\color{blue}{SL: #1}}}
\newcommand*\MHG[1]{{\color{red}{MHG: #1}}}

\newcommand*\edit[1]{{\color{black}{#1}}}

\title{Coupled-cluster pairing models for radicals with strong correlations}

\author{Susi Lehtola}
\orcid{0000-0001-6296-8103}

\affiliation{Chemical Sciences Division, Lawrence Berkeley National Laboratory,
Berkeley, California 94720, United States}

\affiliation{Department of Chemistry, University of Helsinki, P.O. Box 55 (A.
I. Virtasen aukio 1), FI-00014 University of Helsinki, Finland}
\email{susi.lehtola@alumni.helsinki.fi}

\author{Martin Head-Gordon}
\orcid{0000-0002-4309-6669}

\affiliation{Chemical Sciences Division, Lawrence Berkeley National Laboratory,
Berkeley, California 94720, United States}

\affiliation{Department of Chemistry, University of California, Berkeley, California
94720, United States}
\email{mhg@cchem.berkeley.edu}

\begin{abstract}
The pairing hierarchy of perfect pairing (PP), perfect quadruples (PQ) and perfect hextuples (PH) are sparsified coupled-cluster models that are exact in a pairing active
space for 2, 4, and 6 electron clusters, respectively. 
We describe and implement three extensions for radicals.  
First is the trivial generalization that does not correlate radical orbitals.  
The second model (PQr, PHr) includes terms that entangle pair indices and radical indices such that their maximum total number is 2 for PQ and 3 for PH (like their closed-shell versions).  The third family of extended radical models (PPxr, PQxr, and PHxr) include cluster amplitudes that entangle up to 1, 2, and 3 pair indices with up to 1, 2, and 3 radical indices. Notably, PPxr and PQxr are exact for (3e,3o) and (5e,5o), respectively, while still having only $O(N)$ and $O(N^{2}$) amplitudes like their parent models (for $N$ paired electrons).
Orbital optimization is considered for PPxr.  A series of large-scale
numerical tests of these models are presented for spin gaps, and
ionization energies of polyenes and polyenyl
radicals, ranging in size from ethene and allyl radical up to
\ce{C22H24} in full-valence active spaces up to
(122e,122o).
The x\edit{r} 
models perform best.
\end{abstract}
\maketitle
\global\long\def\ERI#1#2{(#1|#2)}

\section{Introduction}
\label{sec:Intro}

Density functional theory\cite{Hohenberg1964_PR_864, Kohn1965_PR_1133} (DFT) is by
far the predominant framework for molecular electronic structure
calculations, because it is computationally inexpensive, and because
it is also sufficiently accurate for most
applications.\cite{Mardirossian2017_MP_2315, Goerigk2017_PCCP_32184} However, although
DFT is formally exact for any system,\cite{Hohenberg1964_PR_864, Kohn1965_PR_1133} in
practice the deficiencies in 
density functional approximations to the exchange-correlation energy make DFT unreliable for systems where
many configurations are important in the wave
function.\cite{Mardirossian2017_MP_2315} Similarly, normal single-reference
wave function methods, such as low-order perturbation theory like
second-order M\o{}ller--Plesset\cite{Moller1934_PR_618} perturbation theory
(MP2), or coupled-cluster (CC) theory with perturbative
triples\cite{Bartlett2012_WIRCMS_126} also do not adequately describe
multi-configurational problems. As these problems arise frequently in
several areas of chemistry, ranging from
biradicaloids\cite{Salem1972_ACIEE_92, Abe2013_CR_7011} to polyradicaloid species such
as multi-metal enzymes\cite{Zhou2018_M_30} and related inorganic molecules
with multiple (or even just one) spin centers,\cite{Marti2008_JCP_14104} the
treatment of strong correlation is a substantial challenge for
electronic structure theory.

It is beyond reasonable limits to fully review all the activity and
excitement that surrounds the strong correlation problem in chemistry
today. However, for our purposes, it is useful to distinguish two
broad approaches to avoiding the exponential cost of the exact
solution of the Schr\"odinger equation. The first class of strong
correlation methods are those that aim for the exact solution to
within some, hopefully acceptable numerical tolerance. Conceptually,
the simplest example of this class are the selected configuration
interaction (CI) methods\cite{Bender1969_PR_23, Langhoff1973_IJQC_999,
  Huron1973_JCP_5745, Buenker1974_TCA_33, Buenker1978_MP_771,
  Evangelisti1983_CP_91, Cimiraglia1987_JCC_39, Illas1991_JCP_1877,
  Harrison1991_JCP_5021, Daudey1993_JCP_1240, Neese2003_JCP_9428,
  Roth2009_PRC_64324, Garniron2018_JCP_64103} that attempt to identify the most important
configurations while discarding the vastly larger set of so-called
``configurational dead-wood''.\cite{Ivanic2001_TCA_339,
  Ivanic2002_TCA_220, Bytautas2009_CP_64} Selected CI methods are
closely related to the full CI quantum Monte-Carlo (FCI-QMC)
approach,\cite{Greer1995_JCP_1821, Greer1998_JCP_181,
  Booth2009_JCP_54106, Booth2010_JCP_174104, Booth2011_JCP_84104,
  Booth2012_JCP_191102, Booth2012_N_365, BenAmor2011_JCP_14101,
  Shepherd2012_JCP_244101, Shepherd2012_PRB_35111,
  Shepherd2012_PRB_81103, Daday2012_JCTC_4441, Giner2013_CJC_879,
  Evangelista2014_JCP_124114, Liu2014_TCA_1481, Liu2016_JCTC_1169,
  Thomas2015_JCP_54108, Giner2015_JCP_44115, Blunt2015_JCP_134117,
  Blunt2015_JCP_184107, Blunt2018_JCP_221101, Blunt2019_JCP_174103,
  Blunt2019_JCTC_3537, Zhang2020_JCTC_9, Weser2022_JCTC_251,
  Weser2023_JCTC_9118}
and are themselves now efficient enough to
handle large systems,\cite{Tubman2016_JCP_44112, Holmes2016_JCTC_3674, Li2018_JCP_214110,
  Tubman2020_JCTC_2139, WilliamsYoung2023_JCP_214109} albeit
still with soft exponential scaling.  Many-body expansions of the
exact energy offer yet another avenue to approach the full CI
energy,\cite{Eriksen2017_JPCL_4633, Zimmerman2017_JCP_104102} as
recently reviewed by \citeit{Eriksen2021_WIRCMS_1525}.  Another
alternative is the density matrix renormalization group (DMRG)
approach\edit{\cite{White1992_PRL_2863, White1993_PRB_10345,
  Schollwoeck2005_RMP_259, Schollwoeck2011_AP_96, Chan2011_ARPC_81,
  Wouters2014_EPJD_272, Szalay2015_IJQC_1342,
  Baiardi2020_JCP_40903, Xu2023_JCTC_4781, Menczer2024_JCTC_8397} and
emerging generalizations thereof,\cite{Marti2010_NJP_103008, Nakatani2013_JCP_134113, Szalay2015_IJQC_1342, Kovyrshin2016_NJP_113001}} which
exploit low-rank wave function separability in a size-extensive way.

A second class of strong correlation methods make model wave functions
that are compact relative either to full CI or to the above
approximations. These methods aim to only capture the most significant
correlations instead of all correlations within a numerical tolerance
as in the first class of approaches. In other words, the methods in
the second class seek a minimal reference wave function for strongly
correlated systems to replace the Hartree--Fock (HF) method of
single-reference theory. We note here that like HF, these methods omit
dynamical correlation, and since dynamical correlation is necessary to
attain quantitative accuracy, in general the methods have to be
corrected in order to become reliable for chemistry; however, in this
work we will focus exclusively on static correlation.

Because the strongest correlations are related to the low-energy
one-particle excitations near the Fermi level, the model wave function
approaches discussed above thus try to solve the Schr\"odinger
equation only in the valence space, via well-defined models. This
straight-away yields complete active space (CAS)
methods,\cite{Roos1980_CP_157, Roos1980_IJQC_175} which solve the
Schr\"odinger equation for some number of active electrons distributed
into some number of active orbitals. As the dimension of the CAS
problem is smaller than that of the untruncated Schr\"odinger
equation, approximations such as selected CI\cite{Smith2017_JCTC_5468, Levine2020_JCTC_2340, Guo2021_JCTC_7545} and DMRG\cite{Zgid2008_JCP_144116, Nakatani2017_JCP_94102, Cheng2022_JPCL_904} can be applied
to the CAS problem more effectively than to the untruncated Schr\"odinger equation.

As originally
suggested by \citeit{Feynman1982_IJTP_467}, quantum computers could
provide accurate solutions to the many-electron problem, with potential for quantum advantage. Quantum phase estimation (QPE) illustrates the potential,\cite{AspuruGuzik2005_S_1704} but requires circuit depths and gate counts that are non-viable on today's NISQ hardware. This has catalyzed the development of more noise-tolerant algorithms, such as the variational quantum eigensolver (VQE)\cite{Peruzzo2014_NC_4213, McClean2016_NJP_23023, Romero2018_QST_14008, Fedorov2022_MT_2} and a host of improved VQE approaches; see \citerefs{Cerezo2021_NRP_625} and \citenum{Tilly2022_PR_1} for recent reviews.
While the promise of quantum computing is bright, present-day and near-term hardware limitations preclude applications that are not readily performed on classical hardware, as is apparent from recent reviews.\cite{Cao2019_CR_10856, Motta2021_WIRCMS_1580, Baiardi2023_C_202300120} 
There is hence ample reason to seek improved classical algorithms until the era of quantum utility for strong correlations is demonstrated.

\edit{Separate
  from methods like selected CI or DMRG that use a non-zero ``working precision'' to \textit{approximately solve} the active space Schr\"odinger equation, one can
  seek well-defined (inexact) polynomial-scaling approximations to the CAS
  problem that are \textit{solved exactly}. CCSD in a valence space is a simple---though quite inaccurate---example of the latter approach. Another} class of such examples
begins with perfect pairing (PP) valence bond (VB)
theory,\cite{Hunt1972_JCP_738, Goddard1978_ARPC_363} and the CC-VB
methods\cite{Small2009_JCP_84103, Small2011_PCCP_19285, Small2012_JCP_114103, 
  Small2014_JCTC_2027, Small2017_JCP_24107, Small2018_JCP_144103, Lee2018_JCP_244121} that approximate
spin-coupled VB (SC-VB).\cite{Gerratt1980_PRSAMPES_525,
  Cooper1988_IRPC_59} by limiting non-orthogonality to within a pair, and using a special cluster expansion inspired by projected Hartree--Fock. A related set of examples are the more general
geminal-based methods,\edit{\cite{Surjan1999__63, Rassolov2002_JCP_5978,
  Johnson2013_CTC_101, Limacher2013_JCTC_1394, Tecmer2014_JPCA_9058,
  Boguslawski2014_PRB_201106,
  Pastorczak2015_PCCP_8622,
  Stein2014_JCP_214113,
  Tecmer2022_PCCP_23026}} as well as minimal matrix product
states.\cite{Larsson2020_JCTC_5057} 
Additionally one can make
coupled-cluster approximations to the active space wave
function,\cite{Krylov1998_JCP_10669, Piecuch1999_JCP_6103,
  Olsen2000_JCP_7140, Piecuch2006_CPL_467, Piecuch2010_MP_2987, Shen2012_JCP_144104, Shen2012_JCTC_4968, 
  Koehn2013_WIRCMS_176,Magoulas_2022} although relatively high order truncations are
required to achieve useful accuracy for problems with strong
correlation character involving more than a single pair of electrons.

To include higher substitutions with lower cost in a coupled-cluster
active space model, a family of generalized perfect pairing (PP)
models has been proposed. 
This hierarchy is an alternative way of truncating the coupled-cluster equations to an active space of $N$ electron pairs, or $2N$ electrons in $2N$ orbitals, commonly denoted as
($2N$e,$2N$o).
To achieve further compactness, the active space in the PP hierarchy is defined as one nominally occupied orbital and one corresponding nominally virtual orbital for each of the $N$ active pairs.\cite{Parkhill2009_JCP_84101,
  Parkhill2010_JCP_24103, Lehtola2016_JCP_134110}
\edit{CC truncation is then defined by exactness (i.e. agreement with FCI) for a given number of electron pairs in the pairing active space: 1 pair in PP,\cite{Hurley1953_PRSAMPES_446, Hunt1972_JCP_738,
  Ukrainskii1977_TMP_816, Goddard1978_ARPC_363, Cullen1996_CP_217,
  Beran2005_JPCA_92} 2 pairs in perfect quadruples\cite{Parkhill2009_JCP_84101} (PQ), and three pairs in perfect hextuples\cite{Parkhill2010_JCP_24103} (PH). Notice that this only requires retaining amplitudes that couple up to the target number of pairs (i.e. 2 at a time for PQ) in excitations that involve up to twice as many electrons (e.g. 4 for PQ).  This powerful truncation using pair locality together with excitation level} should be contrasted to the conventional truncation based on the global excitation level as in CC singles and doubles (CCSD), for example.

Following the aforementioned description, the PP version\cite{Hurley1953_PRSAMPES_446, Hunt1972_JCP_738,
  Ukrainskii1977_TMP_816, Goddard1978_ARPC_363, Cullen1996_CP_217,
  Beran2005_JPCA_92} includes one doubles amplitude per pair of
electrons, and is exact (it agrees with CAS) for a single
pair of electrons, or a set of non-interacting pairs;
the variant of \citeit{Lehtola2016_JCP_134110} can also include two single excitation operators corresponding to the spin-up and spin-down excitations.
The \edit{PQ} model\cite{Parkhill2009_JCP_84101} \edit{already mentioned above} is a truncation of CC
with singles through quadruples (CCSDTQ) with a quadratic number of
amplitudes that yields exactness for 4 electrons in 4 orbitals, or a set of
non-interacting 4-electron-in-4-orbital systems.  Similarly the \edit{PH}
model\cite{Parkhill2010_JCP_24103} is a truncation of CC with single
through hextuple excitations (CCSDTQ56) with a cubic number of
amplitudes that is exact for 6 electrons in 6 orbitals, or a set of non-interacting
6-electrons-in-6-orbital systems.  We have recently reported efficient
implementations of the PQ and PH models including orbital optimization
for problem sizes as large as
(228e,228o).\cite{Lehtola2016_JCP_134110, Lehtola2018_MP_547}

However, generalized PP models have been presented to date only for
molecules with singlet ground states. The purpose of this work is to
explore extending the PP, PQ, and PH hierarchy to the ground state of
systems with a set of $N$ electron pairs, and $R$ radical electrons,
such that \edit{the total spin, $S$, and its $z$ projection, $S_z$ are both equal to $R/2$}. The major question therein is how
to generalize the exactness property to radical electrons. Let
us illustrate this problem with $R=1$ odd electron (\ie{}
$S=1/2$). The simplest alternative is to leave the radical
orbital uncorrelated \edit{(i.e. unentangled)} unlike the active electron pairs, which is
straightforward albeit unappealing, as it will degrade exactness to
the level of just 1 electron for each model---same as
Hartree--Fock. It is more logical, but slightly more complicated, to preserve a
reduced level of accuracy: 1 electron at the PP level (the radical is
unentangled), 3 electrons at the PQ level (\edit{the radical orbital is entangled or correlated 
with each electron pair individually}), and 5 electrons at the PH level (the radical
is entangled with pairs of pairs \edit{in addition}). The most ambitious (and most
complicated) possibility is to provide an \textit{enhanced} level of
accuracy: 3, 5, and 7 electrons at the PP, PQ, and PH levels,
respectively. In this work, we will develop all three possibilities,
as described in detail in \cref{sec:Theory}.

The implementation, described in \cref{sec:Implementation}, makes use of
a computer-based algebra generator for high-order coupled-cluster
theory, as well as our efficient code generator that translates the
sparse tensor contractions into a vectorized form that can be
efficiently evaluated.

We then turn to evaluating the three classes of
pairing models for radicals on some realistic systems.
The computational details are discussed in \cref{sec:compdet}.
Small molecules
are not good choices to examine, because there is no reason to use an
inexact model. We therefore select polyenes and polyenyl radicals as
well as their anions and cations as interesting classes of systems
that have non-trivial electron correlation effects. They have been
widely studied by multireference methods,\cite{Nakayama1998_IJQC_157,
  Kurashige2004_CPL_425} a variety of methods by
\citeit{Bally2000_PCCP_3363}, DMRG,\cite{Hachmann2006_JCP_144101,
  Hu2015_JCTC_3000} SC-VB\cite{Karadakov1994_JACS_2075}, classical
VB,\cite{Luo2004_C_515, Gu2008_JCTC_2101}, CC and related
methods,\cite{Starcke2009_JCP_144311}, scaled opposite-spin
orbital-optimized MP2,\cite{Kurlancheek2012_JCP_54113}, adaptive
CI,\cite{Schriber2017_JCTC_5354} \etc{}. Relevant observables that are
sensitive to the accurate treatment of paired \vs{} radical electrons
include spin gaps, and ionization energies, all
of which are reported on in \cref{sec:results}. Our study concludes with a summary and brief
discussion in \cref{sec:summary}.

\section{Theory}
\label{sec:Theory}

\edit{All the models of the perfect pairing hierarchy are straightforward modifications of standard single-reference coupled-cluster methods. 
The many-electron wave function is still obtained as $|\Psi\rangle = \exp(\hat{T}) |\Psi_0\rangle$, where the excitation operator is still decomposable into single, double, etc operators as $\hat{T} = \hat{T}_1 + \hat{T}_2 + \dots$, and the reference state $|\Psi_0\rangle$ is still a single Slater determinant.
}

\edit{
The central idea in the hierarchy is to include higher-order correlation effects through the minimum number of high-order excitation amplitudes necessary to achieve exactness for a target number of electron pairs. This permits aggressive truncation of the amplitude tensors in an \emph{a priori} manner.
In the following subsections, we will discuss how this truncation is achieved in practice, but the basic idea is to replace the usual truncation with respect to global excitation level (e.g. $\hat{T} = \hat{T}_1+\hat{T}_2$ as in coupled-cluster with singles and doubles, CCSD) with an approach that includes only the miniumum required subset of e.g. the opposite-spin (os) doubles excitation amplitudes 
\begin{equation}
  \hat{T}^\text{os}_2 = \sum_{ijab} t_{i\bar{j}}^{a\bar{b}} a^\dagger_a a^\dagger_{\bar{b}} a_{\bar{j}} a_{i}
\end{equation}
where $i$ and $a$ denote occupied and virtual spin-up orbitals, respectively, and $\bar{j}$ and $\bar{b}$ denote occupied and virtual spin-down orbitals, respectively.
Due to this pair-based truncation of the amplitude tensors, the methods lose some of the invariances of traditional coupled-cluster methods: the energy becomes dependent on occupied-occupied orbital rotations, as well as virtual-virtual orbital rotations, necessitating the orbitals to be optimized.\cite{Lehtola2018_MP_547}
}

\subsection{The PP, PQ, and PH models}
\label{ssec:parent_models}

\begin{figure}[t]
\begin{centering}
\includegraphics[width=6cm]{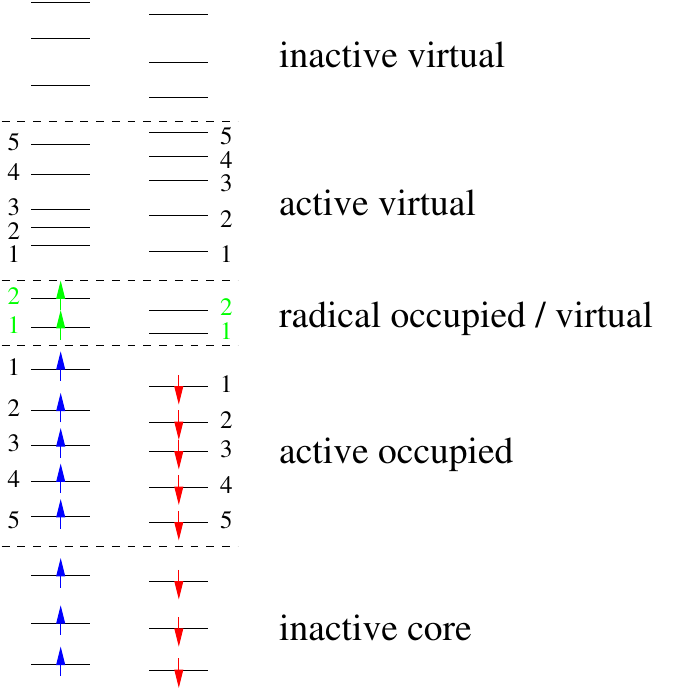}
\par\end{centering}
\caption{Illustration of the division of the pairing orbital space, including the new
  unpaired radical class of electrons and
  orbitals.
  The example features a (12e,12o) problem for a triplet state, in which the active space is split up into a traditional (10e,10o) perfect pairing active space and a (2e,2o) radical subspace. The radical pairing models are formed by selecting suitable active space truncations of the coupled cluster models. \label{fig:Division-of-the}}
\end{figure}

As already mentioned in \cref{sec:Intro}, the PP, PQ, and PH models all use a pairing active space, in which
each occupied orbital, $i$, has a single correlating virtual orbital,
$i^{*}$, as illustrated in \cref{fig:Division-of-the}. Let us first consider the CC version\cite{Cullen1996_CP_217,
  Beran2005_JPCA_92} of the PP model, which is exact for isolated
pairs. Here, we thus consider single
excitations from the occupied alpha and beta orbitals $i$ and
$\bar{i}$ to the corresponding virtuals $i^{*}$ and $\bar{i}^{*}$,
respectively, and the double excitation from the occupied alpha and
beta orbitals $i$ and $\bar{i}$ to the corresponding virtuals $i^{*}$
and $\bar{i}^{*}$, the overbar denoting beta spin. Although the
singles excitation amplitudes are not traditionally considered within PP, we
have shown them to be important even at optimal
orbitals\cite{Lehtola2018_MP_547} as previously suggested by
\citeit{Koehn2005_JCP_84116}.

To illustrate the truncation, we will consider the opposite-spin block
of the double excitation amplitudes tensor, which is approximated in
PP as\cite{Lehtola2016_JCP_134110}
\edit{
\begin{align}
\left(t_\textrm{PP}\right)_{i\bar{j}}^{a\bar{b}}= & \sum_{p=1}^{N}t_{p\bar{p}}^{p^*\bar{p}^*}\delta_{ip}\delta_{\bar{j}\bar{p}}\delta_{ap^*}\delta_{\bar{b}\bar{p}^*}, \label{eq:pp}
\end{align}
where $t_{p\bar{p}}^{p^*\bar{p}^*}$ is the PP opposite-spin double excitation amplitude for pair $p$.
There is no contribution from the same-spin doubles due to fermionic symmetry.}
The $\mathcal{O}(N^4)$ amplitudes of CCSD are thereby reduced to $N$ non-zero
values in PP \edit{($3N$ if the single excitations $t_{p}^{p^*}$ and $t_{\bar{p}}^{\bar{p}^*}$ are also included in the calculation)}, even though exactness within (2e,2o) active spaces and
size-consistency\cite{Pople1976_IJQC_1} are unaffected.  In the PQ model,\cite{Parkhill2009_JCP_84101}
all CC terms coupling two pairs are retained to ensure accuracy for 4
electrons in 4 orbitals (\ie{} 2 pairs) while maintaining the
size-consistency\cite{Pople1976_IJQC_1} of CC. Thus, quadratic subsets (``2P'') of the
singles, doubles, triples and quadruples amplitudes are retained in PQ such
that
\begin{equation}
{{\bf{t}}_{{\rm{PQ}}}} = {{\bf{t}}_{{\rm{PP}}}} + {{\bf{t}}_{2{\rm{P}}}}
\end{equation}
For instance, to compare with PP and \cref{eq:pp},
${{\bf{t}}_{2{\rm{P}}}}$ for the opposite-spin block of the double
excitation amplitudes tensor is defined as\cite{Lehtola2016_JCP_134110}
\edit{
\begin{widetext}
\begin{align}
\left(t_\textrm{2P}\right)_{i\bar{j}}^{a\bar{b}} = & \sum_{\substack{p,q=1\\
p\neq q
}
}^{N} t_{p\bar{p}}^{p^* \bar{q}^*}\delta_{ip}\delta_{\bar{j}\bar{p}}\delta_{ap^*}\delta_{\bar{b}\bar{q}^*}+\sum_{\substack{p,q=1\\
p\neq q
}
}^{N} t_{p\bar{p}}^{q^*\bar{p}^*}\delta_{ip}\delta_{\bar{j}\bar{p}}\delta_{aq^*}\delta_{\bar{b}\bar{p}^*}+\sum_{\substack{p,q=1\\
p\neq q
}
}^{N} t_{p\bar{q}}^{p^*\bar{p}^*}\delta_{ip}\delta_{\bar{j}\bar{q}}\delta_{ap^*}\delta_{\bar{b}\bar{p}^*}\nonumber \\
+ & \sum_{\substack{p,q=1\\
p\neq q
}
}^{N} t_{q\bar{p}}^{p^*\bar{p}^*}\delta_{iq}\delta_{\bar{j} \bar{p}}\delta_{ap^*}\delta_{\bar{b}\bar{p}^*}+\sum_{\substack{p,q=1\\
p\neq q
}
}^{N} t_{p\bar{q}}^{p^*\bar{q}^*}\delta_{ip}\delta_{\bar{j} \bar{q}}\delta_{ap^*}\delta_{\bar{b}\bar{q}^*}+\sum_{\substack{p,q=1\\
p\neq q
}
}^{N} t_{q\bar{p}}^{p^*q^*}\delta_{iq}\delta_{\bar{j} \bar{p}}\delta_{ap^*}\delta_{\bar{b} \bar{q}^*}+\sum_{\substack{p,q=1\\
p\neq q
}
}^{N} t_{p\bar{p}}^{q^* \bar{q}^*}\delta_{ip}\delta_{\bar{j}\bar{p}}\delta_{aq^*}\delta_{\bar{b} \bar{q}^*}. \label{eq:pq}
\end{align}
\end{widetext}
} \noindent There are corresponding ${{\bf{t}}_{2{\rm{P}}}}$ terms for the same
spin doubles, as well as the various spin blocks of the single,
triple, and quadruple substitutions. Overall, the $O(N^8)$ amplitudes
of active space CCSDTQ are thereby reduced to $O(N^2)$ non-zero values
in PQ, while exactness within (4e,4o) active spaces and
size-consistency\cite{Pople1976_IJQC_1} are unaffected.

To define the PH model\cite{Parkhill2010_JCP_24103} which is exact for 3 pairs
or (6e,6o) active spaces, we introduce a similar equation that
contains the additional terms beyond PQ that couple 3 electron pairs
(``3P''). This defines the sparsity pattern of the amplitude
tensors in the PH method as 
\begin{equation}
{{\bf{t}}_{{\rm{PH}}}} = {{\bf{t}}_{{\rm{PQ}}}} +
{{\bf{t}}_{3{\rm{P}}}}
\end{equation}
Again, as an example, ${{\bf{t}}_{3{\rm{P}}}}$ for the the
opposite-spin block of the double excitation amplitudes tensor is
defined as:
\edit{
\begin{widetext}
\begin{align}
\left(t_\textrm{3P}\right)_{i\bar{j}}^{a\bar{b}} =  & \sum_{\substack{p,q,r=1\\
p\neq q\neq r
}
}^{N}t_{p\bar{p}}^{q^*\bar{r}^*}\delta_{ip}\delta_{\bar{j} \bar{p}}\delta_{aq^*}\delta_{\bar{b} \bar{r}^*}+\sum_{\substack{p,q,r=1\\
p\neq q\neq r
}
}^{N}t_{q \bar{r}}^{p^* \bar{p}^* }\delta_{iq}\delta_{\bar{j}\bar{r}}\delta_{ap^*}\delta_{\bar{b} \bar{p}^*}+\sum_{\substack{p,q,r=1\\
p\neq q\neq r
}
}^{N} t_{p\bar{q}}^{p^* \bar{r}^*}\delta_{ip}\delta_{\bar{j} \bar{q}}\delta_{ap^*}\delta_{\bar{b} \bar{r}^*}\label{eq:ph}\\
+ & \sum_{\substack{p,q,r=1\\
p\neq q\neq r
}
}^{N} t_{p\bar{q}}^{r^* \bar{p}^*}\delta_{ip}\delta_{\bar{j} \bar{q}}\delta_{ar^*}\delta_{\bar{b}\bar{p}^*}+\sum_{\substack{p,q,r=1\\
p\neq q\neq r
}
}^{N} t_{q\bar{p}}^{p^* \bar{r}^*}\delta_{iq}\delta_{\bar{j} \bar{p}}\delta_{ap^*}\delta_{\bar{b} \bar{r}^*}+\sum_{\substack{p,q,r=1\\
p\neq q\neq r
}
}^{N}t_{q\bar{p}}^{r^* \bar{p}^*}\delta_{iq}\delta_{\bar{j} \bar{p}}\delta_{ar^*}\delta_{\bar{b} \bar{p}^*}
\nonumber
\end{align}
\end{widetext}
} \noindent To ensure exactness for 3 pairs, PH has a large number of other
sparse tensors with a cubic number of amplitudes (from 3 pair indices)
all the way through hextuple substitutions. Overall, the $O(N^{12})$
amplitudes of active space CCSDTQ56 are thereby reduced to $O(N^3)$
non-zero values in PH, while exactness within (6e,6o) active spaces
and size-consistency\cite{Pople1976_IJQC_1} are again unaffected.

\subsection{The PPr, PQr and PHr models}
\label{ssec:r_models}

As discussed in \citeref{Lehtola2016_JCP_134110}, the representation in
\crefrange{eq:pp}{eq:ph} contains an inherent assumption that the active
orbital blocks are of equal size. Let us now remove that assumption
and instead allow the presence of a set of $R$ radical orbitals, each
accompanied by a radical electron. The resulting nominal ground state
determinant is illustrated in \cref{fig:Division-of-the}. How should the
opposite-spin block of the double excitation amplitudes tensor (and
its other spin blocks and other relevant tensors) be approximated now?
As foreshadowed in the Introduction, there are three
possibilities. The first possibility is the trivial generalization
that leaves the definitions of the PP, PQ, and PH sparse amplitude
tensors unmodified in their form discussed above for
radicals. Therefore, no radical orbitals enter these models, and
obviously, no modifications to the equations are required; however,
the radical electrons are not correlated and the models lose their
exactness properties.

Any other alternative must include additional terms involving, that is,
entangling, the radical orbitals and electrons with the paired
electrons. The simplest logical way to do this is to include
additional terms that have no more labels (either pair or radical
labels) than the base model, which defines our second class of
candidate open shell models. We shall denote such models as PPr, PQr,
and PHr, where the suffix ``r'' denotes the fact that they are
generalizations for molecules with radical electrons.  We can write
the following relations to define these models:
\begin{align}
{{\bf{t}}_{{\rm{PPr}}}} & = {{\bf{t}}_{{\rm{PP}}}}
\label{eq:PPr} \\
{{\bf{t}}_{{\rm{PQr}}}} & = {{\bf{t}}_{{\rm{PPr}}}} + {{\bf{t}}_{2{\rm{P}}}} + {{\bf{t}}_{2{\rm{r}}}}
\label{eq:PQr} \\
{{\bf{t}}_{{\rm{PHr}}}} & = {{\bf{t}}_{{\rm{PQr}}}} + {{\bf{t}}_{3{\rm{P}}}} + {{\bf{t}}_{3{\rm{r}}}}
\label{eq:PHr}
\end{align}
The equivalence of PP and PPr follows from the fact that any
correlation between electron pairs and radicals (r) must involve a
minimum of two indices: one for the radical label and one for the pair
label. This minimal coupling defines the additional
${{\bf{t}}_{2{\rm{r}}}}$ terms that augment the PQr model relative to
PQ. Likewise, ${{\bf{t}}_{3{\rm{r}}}}$ is defined by amplitudes that
entangle 3 different indices, at least one of which is a radical
index, and these terms augment the PHr model relative to PH.

For concrete illustration of ${{\bf{t}}_{2{\rm{r}}}}$ and
${{\bf{t}}_{3{\rm{r}}}}$, let us again focus on the opposite-spin
block of the double excitation amplitudes tensor. At the 2r level, one
pair index and one radical index can now entangle so that 3 new terms
are retained: \edit{ \begin{widetext}
\begin{align}
\left(t_\textrm{2r}\right)_{i\bar{j}}^{a\bar{b}} =
 \sum_{p=1}^{N}\sum_{x=1}^{R}t_{x\bar{p}}^{p^* \bar{p}^*}\delta_{ix}\delta_{\bar{j} \bar{p}}\delta_{ap^*}\delta_{\bar{b} \bar{p}^*}+\sum_{p=1}^{N}\sum_{x=1}^{R}t_{p\bar{p}}^{p\bar{x}}\delta_{ip}\delta_{\bar{j}\bar{p}}\delta_{ap^*}\delta_{\bar{b}\bar{x}}
+ \sum_{p=1}^{N}\sum_{x=1}^{R}t_{xp}^{p^*\bar{x}}\delta_{ix}\delta_{\bar{j} \bar{p}^*}\delta_{ap^*}\delta_{\bar{b}\bar{x}}.
\label{eq:2r}
\end{align}
\end{widetext}
} \noindent As a reminder, these additional terms denote the excitations
$(x,\bar{i})\to(i^{*},\bar{i}^{*})$ (radical electron to paired virtual), $(i,\bar{i})\to(i^{*},\bar{x})$ (paired electron to radical virtual),
and $(x,\bar{i})\to(i^{*},\bar{x})$ (radical electron to paired virtual, and paired electron to radical virtual). At the level of 3 entangled
indices, two pair indices and one radical index, or one pair index and
two radical indices can be entangled, giving rise to the following
additional terms
\edit{
\begin{widetext}
\begin{align}
  \left(t_{\textrm{3r}}\right)_{i\bar{j}}^{a\bar{b}}&=\sum_{\substack{p,q=1\\
      p\neq q
    }
  }^{N}\sum_{x=1}^{R} t_{x\bar{q}}^{p^* \bar{p}^*}\delta_{ix}\delta_{\bar{j} \bar{q}}\delta_{ap^*}\delta_{\bar{b} \bar{p}^*}+\sum_{p=1}^{N}\sum_{\substack{x,y=1\\
      x\neq y
    }
  }^{R}t_{x\bar{p}}^{p^*\bar{y}^*}\delta_{ix}\delta_{\bar{j} \bar{p}}\delta_{ap^*}\delta_{\bar{b}\bar{y}^*}+\sum_{\substack{p,q=1\\
      p\neq q
    }
  }^{N}\sum_{x=1}^{R}t_{x\bar{p}}^{p\bar{q}}\delta_{ix}\delta_{\bar{j}\bar{p}}\delta_{ap^*}\delta_{\bar{b}\bar{q}^*} \nonumber \\&+\sum_{\substack{p,q=1\\
      p\neq q
    }
  }^{N}\sum_{x=1}^{R}t_{x\bar{q}}^{p^* \bar{x}}\delta_{ix}\delta_{\bar{j} \bar{q}}\delta_{ap^*}\delta_{\bar{b} \bar{x}}+\sum_{\substack{p,q=1\\
      p\neq q
    }
  }^{N}\sum_{x=1}^{R}t_{x\bar{p}}^{q^*\bar{p}^*}\delta_{ix}\delta_{\bar{j} \bar{p}}\delta_{aq^*}\delta_{\bar{b} \bar{p}^*}+\sum_{\substack{p,q=1\\
      p\neq q
    }
  }^{N}\sum_{x=1}^{R}t_{p\bar{q}}^{p^* \bar{x}}\delta_{ip}\delta_{\bar{j} \bar{q}}\delta_{ap^*}\delta_{\bar{b}\bar{x}} \nonumber \\&+\sum_{\substack{p,q=1\\
      p\neq q
    }
  }^{N}\sum_{x=1}^{R}t_{q\bar{p}}^{p^* \bar{x}}\delta_{iq}\delta_{\bar{j} \bar{p}}\delta_{ap^*}\delta_{\bar{b} \bar{x}}+\sum_{\substack{p,q=1\\
      p\neq q
    }
  }^{N}\sum_{x=1}^{R}t_{p\bar{p}}^{q^* \bar{x}}\delta_{ip}\delta_{\bar{j} \bar{p}}\delta_{aq^*}\delta_{\bar{b} \bar{x}}
   \label{eq:3r}
\end{align}
\end{widetext}
} \noindent that correspond to the $(x,\bar{j})\to(i^{*},\bar{i}^{*})$,
$(x,\bar{i})\to(i^{*},\bar{y})$,
$(x,\bar{i})\to(i^{*},\bar{j}^{*})$,
$(x,\bar{j})\to(i^{*},\bar{x})$,
$(x,\bar{i})\to(i^{*},\bar{j}^{*})$,
$(i,\bar{j})\to(i^{*},\bar{x})$,
$(j,\bar{i})\to(i^{*},\bar{x})$, and
$(i,\bar{i})\to(j^{*},\bar{x})$ excitations, $i$ and $j$ denoting
pair indices and $x$ and $y$ denoting radical indices.

Let us briefly consider the exactness properties of the PPr, PQr and
PHr models. \edit{As a reminder, by exactness, we mean agreement with FCI in the same space of active orbitals, since we seek to approximate CASSCF.} For PPr, exactness is maintained for isolated pairs of
electrons (not just singlets but also triplets in a trivial fashion, because there are no
empty $\alpha$ orbitals), as well as isolated radicals. However, a
three-electron doublet fragment cannot be exact, although a quartet
three-electron fragment is trivially exact. Moving on to PQr, the
structure of the additional terms makes it clear that PQr will be
exact for doublet three-electron systems, and non-interacting clusters
of three-electron and four-electron systems, including triplet
fragments, and subsets thereof. Finally, it is evident that PHr will
be exact for five-electron doublet or higher-spin systems, and
non-interacting clusters of five- and six-electron systems, and
subsets thereof. In summary, for doublet systems, this hierarchy of
radical models is exact for $(2n-1)$ electron systems ($n=1,2,3$ for
PPr, PQr, and PHr) in the pairing active space, and exact for
$2n$-electron singlet and triplet multiplicities.


\subsection{The PPxr, PQxr and PHxr models}
\label{ssec:xr_models}

There is another possible generalization of the pairing models to
radical systems that has more desirable exactness properties than the
models introduced above.  Let us recall that the
${{\bf{t}}_{2{\rm{r}}}}$ terms were added to the base PQ model to
define the PQr model via \cref{eq:PQr}, while the PPr model was
unchanged from PP itself. We can instead augment the PP model itself
with the ${{\bf{t}}_{2{\rm{r}}}}$ to yield a version of PP with
extended radical (xr) correlations:
\begin{equation}
{{\bf{t}}_{{\rm{PPxr}}}} = {{\bf{t}}_{{\rm{PP}}}} + {{\bf{t}}_{2{\rm{r}}}}
\end{equation}
From the previous subsection, we recall that the
${{\bf{t}}_{2{\rm{r}}}}$ terms entangle radicals with pairs (e.g. as
illustrated in \cref{eq:2r} for the opposite-spin block of the double
excitation amplitudes). By definition there must be at least one
radical index, which means that there is at most one pair index in
each amplitude in ${{\bf{t}}_{2{\rm{r}}}}$. Thus inclusion of these
one-pair index radical correlations in the PPxr model is consistent
with the inclusion of only one-pair correlations in PP
itself. Furthermore, if we are interested in the doublet manifold, we
note that $R=1$ for all doublet states and therefore the number of
retained amplitudes in ${{\bf{t}}_{2{\rm{r}}}}$ is only $3N$, similar
to the $N$ amplitudes contained in ${{\bf{t}}_{{\rm{PP}}}}$.

In the same spirit, it would also be possible to include terms of the
type $(x,\bar{i}) \to (i^{*},\bar{y})$ in an extended PP model with
$O(N)$ amplitudes; however, for simplicity, we will restrict the
number of radical indices similarly to the pair indices, allowing up
to 2 radical indices for PQxr and up to 3 radical indices for PHxr.
Now, inductive generalization suggests that PQxr and PHxr models
should be defined as follows:
\begin{align}
{{\bf{t}}_{{\rm{PQxr}}}} & = {{\bf{t}}_{{\rm{PPxr}}}} + {{\bf{t}}_{2{\rm{P}}}} + {{\bf{t}}_{3{\rm{r}}}} + {{\bf{t}}_{4{\rm{r}}}} \label{eq:PQxr} \\
{{\bf{t}}_{{\rm{PHxr}}}} & = {{\bf{t}}_{{\rm{PQxr}}}} + {{\bf{t}}_{3{\rm{P}}}} + {{\bf{t}}_{5{\rm{r}}}} + {{\bf{t}}_{6{\rm{r}}}} \label{eq:PHxr}
\end{align}
In direct analogy to the discussion above for PPxr, the
${{\bf{t}}_{3{\rm{r}}}}$ and ${{\bf{t}}_{4{\rm{r}}}}$ terms now
incorporated into PQxr include no more than two pair indices, just
like the parent PQ model itself; however, up to two radical indices
can be included, leading to a maximum of 4 indices in the
subtensor. Likewise the ${{\bf{t}}_{5{\rm{r}}}}$ and
${{\bf{t}}_{6{\rm{r}}}}$ terms incorporated into PHxr include
correlations between radical and pair indices so that no more than
three pair indices can appear just like in the parent closed-shell PH
model, but in addition up to three radical indices can be included,
yielding up to six indices in total.

Let us next consider the exactness properties of the PPxr, PQxr and
PHxr models \edit{(as before, meaning exactness against FCI in the same active space of pairing and radical orbitals)}. For PPxr, exactness is of course maintained for isolated
pairs of electrons (both singlets and triplets) as discussed above in \cref{ssec:r_models} for PPr which is a subset of PPxr. 
In contrast to PPr, three-electron
doublet fragments are also exact in PPxr, due to inclusion of the
${{\bf{t}}_{2{\rm{r}}}}$ terms. Note that triple substitutions are not
necessary for (3e,3o) exactness as there is only one $\alpha$ virtual,
and only one $\beta$ occupied. Therefore, like PP and PPr, PPxr is a subset of
CCSD. It is interesting to note that PPxr is also exact for the (4e,4o)
triplet, the (5e, 5o) quartet, \etc{}, each of which also requires
only double substitutions.

Moving on to PQxr, the structure of the additional terms mandates that
the model is exact for doublet 5 electron systems, while PQr was only exact for doublet 3 electron systems. Quintuple
substitutions are not necessary for (5e,5o) exactness, so PQxr is a
subset of active-space CCSDTQ like PQ and PQr. Similarly PQxr is exact
for non-interacting clusters of five electrons, and other
four-electron clusters due to its size-consistency. 
PQxr is also exact for six-electron
triplets, seven-electron quartets \etc{}, while still being a subset
of CCSDTQ like PQr. Finally, PHxr is exact for seven-electron doublet (or
higher-spin) systems, and non-interacting clusters of seven- and
six-electron systems, and subsets thereof, all while still being a
subset of CCSDTQ56 like its corresponding closed-shell PH parent
model and the PHr model.

This kind of truncation then yields (3e,3o) exactness for PPxr, the
(6e,6o) triplet for PQxr, and the (9e,9o) quintet for PHxr, and
subsets thereof.

\begin{table*}
\begin{centering}
\begin{tabular}{lc}
 & $t/\lambda$\tabularnewline
\hline
\hline
PP (and PPr) & $3N$\tabularnewline
PPxr & $3N+5NR$\tabularnewline
PQ & $3N+16N^{2}$\tabularnewline
PQr & $3N+5NR+16N^{2}$\tabularnewline
PQxr & $3N+5NR+NR^{2}+16N^{2}+27N^{2}R+20N^{2}R^{2}$\tabularnewline
 & \tabularnewline
 & $\gamma$\tabularnewline
\hline
\hline
PP (and PPr) & $6N$\tabularnewline
PPxr & $2R+6N+6NR$\tabularnewline
PQ & $6N+6N^{2}$\tabularnewline
PQr and PQxr & $2R+2R^{2}+6N+6NR+6N^{2}$\tabularnewline
 & \tabularnewline
 & $\Gamma$\tabularnewline
\hline
\hline
PP (and PPr) & $13N$\tabularnewline
PPxr & $R+13N+55NR$\tabularnewline
PQ & $13N+105N^{2}$\tabularnewline
PQr & $R+5R^{2}+13N+55NR+105N^{2}$\tabularnewline
PQxr & $R+9R^{2}+13N+57NR+49NR^{2}+105N^{2}+117N^{2}R+31N^{2}R^{2}$\tabularnewline
\end{tabular}
\par\end{centering}
\caption{Storage costs for the $t$ and $\lambda$ amplitudes, and the
  one- and two-electron density matrices $\gamma$ and $\Gamma$,
  respectively, that appear in some of the models considered in the
  present work, in terms of the number of electron pairs, $N$, and the
  number of unpaired radical electrons, $R$.
\label{tab:Storage-costs-for}}
\end{table*}

\section{Implementation}
\label{sec:Implementation}

The implementation of the new models is done using the same framework
and automatic code generator as in \citeref{Lehtola2016_JCP_134110},
which proceeds briefly as follows. Starting from the standard CC
equations in spin-orbital form that have been generated with a
computer algebra system,\cite{Parkhill2010_MP_513} the spin is
integrated out to obtain the equations in terms of the various spin
blocks of the (de-)excitation tensors, the integrals and the density
matrices. 
Next, the decompositions for the spin blocks, exemplified by
\cref{eq:pp,eq:pq,eq:2r,eq:3r} \etc{}, are generated, and
summations over the Kronecker symbols are performed in the CC
equations. This yields the equations for the pairing models in terms
of the dense subtensors only. \edit{As in our previous works, the $\lambda$ amplitudes, the one- and two-electron density matrices, and the integrals are truncated analogously to the $t$ amplitudes.}
Finally, the generator writes out C++ code that implements the
equations. 
The resulting storage costs for the $t$
excitation amplitudes, the $\lambda$ de-excitation amplitudes, and the
one- and two-electron density matrices are given in
\cref{tab:Storage-costs-for}.

Due to the significant number of additional terms that arise from the
presence of radical orbitals, in the present work we will only
consider the full PPxr and PQxr models, and the CCSDTQ subset of PHr
and PHxr. The code generator at its present stage of development
cannot feasibly form the terms needed for the full models, as the
number of possible labelings grows exponentially in the order of the
original tensors, and the tensor product pairing algorithm is cubic
scaling in the number of labelings. A complete refactor (rewrite) of the
proof-of-concept pairing code generator is expected to make generating
the full PHxr model tractable, but this undertaking is beyond the scope
of the present proof-of-concept work. The partial implementation of the radical PH
models will be denoted as PHrQ or and PHxrQ to indicate that only
terms through quadruples are retained.
The partial implementation of the PH model is analogously denoted as PHQ.

The radical models are generated with two-pair, three-pair, and
four-pair intermediates for PP, PQ and PH level, respectively. That
choice is in accordance to the truncation scheme chosen in
\cref{sec:Theory}. For instance the two-pair intermediates of PPxr may
have up to two pair labels as well as up to two radical orbital
labels.

\section{Computational Details}
\label{sec:compdet}

\begin{figure*}
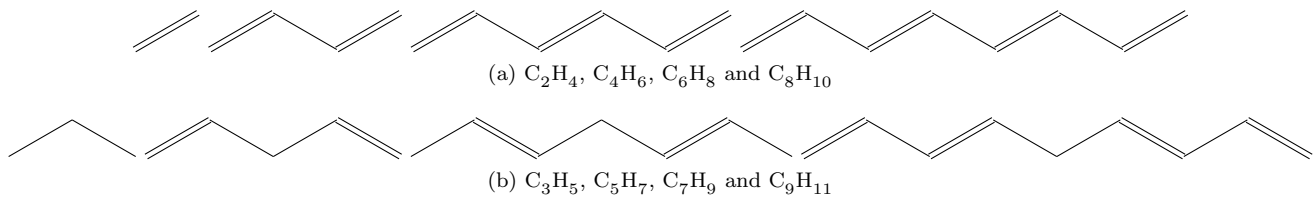

  \subfloat[\ce{C2H4}, \ce{C4H6}, \ce{C6H8} and \ce{C8H10}]{
    \chemfig{[:30]=}

    \chemfig{[:30]=-[:-30]=[:30]}

    \chemfig{[:30]=-[:-30]=-[:-30]=[:30]}

    \chemfig{[:30]=-[:-30]=-[:-30]=-[:-30]=[:30]}
  }

  \subfloat[\ce{C3H5}, \ce{C5H7}, \ce{C7H9} and \ce{C9H11}]{
    \chemfig{[:30]--[:-30]}

    \chemfig{[:30]=-[:-30]-=[:-30]}

    \chemfig{[:30]-=[:-30]--[:-30]=-[:-30]}

    \chemfig{[:30]=-[:-30]=-[:-30]-=[:-30]-=[:-30]}
  }

  \caption{Illustration of the used model geometries for the polyenes and
    polyenyls; larger molecules are obtained by continuation of the
    series. The single and double lines denote single and double bonds, whose bond lengths are given in the main text. \label{fig:molstruct}}
\end{figure*}

In the present work, we study polyene and polyenyl molecules at fixed
model geometries. Following \citeit{Hachmann2006_JCP_144101} the
polyene geometries are extracted from density functional calculations
extrapolated to the bulk limit,\cite{Catalan2004_JCP_1864} yielding the
parameters R(\ce{C\bond{=}C}) = 1.3693 \AA , R(\ce{C\bond{-}C}) =
1.4244 \AA, and R(\ce{C\bond{-}H}) = 1.0820 \AA, with all bond angles
equalling $120^{\circ}$ and the molecules being planar. In the present
work, for simplicity, the same parameters are also used for the
polyenyls. 
To account for the polyenyls' radical character, the polyenyls are constructed to contain two single bonds at the middle of the molecule to host the radical electron.
The molecular structures
are illustrated in \cref{fig:molstruct}. Both polyenes and polyenyls
possess $C_{s}$ symmetry; symmetry restrictions were, however, not
imposed in the calculations.

The automatically generated pairing models have been interfaced with
the ERKALE program,\cite{Lehtola2012_JCC_1572, Lehtola2018__a} which
is used to generate the integrals and perform the orbital optimization \edit{of the orthonormal molecular orbitals}
with the geometric direct minimization (GDM)
method\cite{VanVoorhis2002_MP_1713, VanVoorhis2002_JCP_9190,
  Dunietz2002_JTCC_255} in which the descent direction is determined
by a Broyden--Fletcher--Goldfarb--Shanno (BFGS)
algorithm\cite{Nocedal1999__} preconditioned with diagonal second
derivatives, and the optimization is continued until the norm of the orbital gradient
satisfies $\left\Vert
\partial\mathcal{L}/\partial\theta_{pq}\right\Vert <10^{-5}$.


The cc-pVDZ basis set\cite{Dunning1989_JCP_1007} is employed in all
calculations, and the pairing active space is used for all valence
electrons.  That is, a full valence space of (($5n$+2)e,($5n$+2)o) is
used for for the \ce{C_{n}H_{n+2}} neutral species, whereas for
cationic species the active space is reduced to a quasi-full-valence
(($5n$+1)e,($5n$+1)o) active space. Thus, the active spaces range from
(11e,11o) for \ce{C2H4+} to (122e,122o) for \ce{C22H24}. 
Note that the size of the active space is the same regardless of the studied spin state, but the composition of the active space does depend on the number of unpaired electrons.
Single-point
calculations are run with the PP, PPr, PPxr, PQ, PQr, PQxr, PHQ, PHrQ, and PHxrQ models up through \ce{C16H18}
\ie{} active spaces up to (82e,82o).
All the single-point calculations include single
excitations within the active space, allowing the orbitals to relax
within the active space.

As in our previous work,\cite{Lehtola2018_MP_547} the calculations are
initialized using ROHF electron densities. A suitable initial guess is
generated by localizing the occupied
orbitals\cite{Lehtola2013_JCTC_5365} using the generalized
Pipek--Mezey criterion\cite{Lehtola2014_JCTC_642} using Becke
charges. The generalized Pipek--Mezey localization is run separately
for the paired active space and for the radical orbitals, after which
corresponding virtual orbitals are generated for the paired active
orbitals using the Sano guess.\cite{Sano2000_JMST_177,
  VanVoorhis2002_JCP_9190} Two-electron integrals and Fock matrices
are evaluated using the Cholesky decomposition
method,\cite{Beebe1977_IJQC_683, Pedersen2023_WIRCMS_1692} following
the approach of \citeref{Koch2003_JCP_9481} with a $10^{-10}$ integral
screening threshold and a $10^{-9}$ threshold for the Cholesky
procedure itself.

\edit{Although we implemented orbital optimization for all the models considered herein, orbital optimization for the higher models was found to be difficult, which we tentatively attribute to the models becoming more exact: going up the hierarchy, the models become less and less sensitive to the employed orbitals, and possibly also introduce additional local minima.}
Since orbital optimization is \edit{anyways} only practical for large calculations
with the basic PP model and its PPxr extension, we must make a choice
of which model to use for assessments of energy differences such as
spin gaps and ionization energies. To ensure spin-pure states, we
choose to use restricted orbitals (RO), and to include some correlation
effects associated with the radical orbitals we use the PPxr model to
optimize the orbitals. Thus all single-point energy differences
reported in the following section employ RO-PPxr optimized orbitals,
together with the standard geometries discussed above, and the cc-pVDZ
basis set. 
\edit{The orbital optimizations within the PPxr model were found to converge within a few dozen up to a few hundred iterations.
We note again that the use of PPxr orbitals is justified, since the choice of the orbitals becomes less and less important going up in the pairing hierarchy.
Single-point calculations with these orbitals are able to approach the exact solution within the employed orbital active space, but the PPxr model should be quite good at identifying which orbitals are significantly correlated.
}

\edit{In the following section, we present vertical singlet-triplet ($\Delta E^\text{ST}$) and doublet-quartet ($\Delta E^\text{DQ}$) gaps as well as ionization energies ($\Delta E^\text{IE}$).
All of these data are computed with the $\Delta$SCF methodology by subtraction of ground-state energies for the corresponding species: $\Delta E^\text{ST} = E(\text{triplet})-E(\text{singlet})$, $\Delta E^\text{DQ} = E(\text{quartet})-E(\text{doublet})$, and $\Delta E^\text{IE} = E(\text{cation})-E(\text{neutral})$.}

\section{Results}
\label{sec:results}

\subsection{Singlet-triplet gaps of polyenes}
Vertical singlet-triplet (ST) gaps for the polyenes are given in
\cref{tab:stgaps-xpp}. Calculated ST gaps are an interesting measure of
balance in the treatment of differential correlation effects when the
pairing of two electrons is disrupted. In our context, the PP, PQ and
PHQ models \textit{clearly undercorrelate} the radical electrons of
the triplet. This trend is expected: since the radical orbitals do not
enter any amplitudes, the radical orbitals are not correlated at
all. The trend is manifested in the \textit{strongly increasing} ST
gaps for the PP, PQ, PHQ sequence, where pair correlations are
increasingly complete, whilst the radical orbitals remain
uncorrelated.

The r and xr models may then offer potentially more balanced
treatments of two electrons that are paired versus unpaired. This
improved balance is evident in the results of \cref{tab:stgaps-xpp}, as
much smaller changes occur across the PP, PQ, PH sequence for the r
and xr models. While our calculations use model geometries, and are
performed only in the quasi-full-valence active space, it is evident
that PPxr, and particularly PQxr and PHxrQ compare very well with both
experimental values, and with multireference M\o{}ller--Plesset (MRMP)
reference values. The dramatic improvement of PPxr relative to PP is
noteworthy.
It is also
noteworthy that the values for PQxr agree much better with the
reference values for the longer chains than the ones for PQr. Finally,
only very small changes are seen moving from PQxr to PHxrQ, which is
the most complete method implemented here. All these observations
suggest that the xr models are significantly better balanced than the
r models.

\begin{table*}
\begin{threeparttable}[t]
\setlength{\tabcolsep}{0.5em} 
\begin{tabular}{c|ccc|cc|ccc|c|c}
\toprule
$n_{C}$ & PP & PQ & PHQ & PQr  & PHrQ & PPxr & PQxr & PHxrQ   & MRMP\tnote{a} & expt.\tnote{b}\tabularnewline
\hline
2  & 3.971 & 4.509 & 4.585 & 4.387 & 4.399 & 3.869 & 4.295 & 4.375 &      & 4.32-4.36$^b$\tabularnewline
4  & 3.554 & 4.327 & 4.516 & 3.455 & 3.322 & 2.805 & 3.106 & 3.210 & 3.20 & 3.22$^c$\tabularnewline
6  & 3.361 & 4.327 & 4.576 & 3.268 & 2.820 & 2.527 & 2.511 & 2.626 & 2.40 & 2.58-2.61$^d$\tabularnewline
8  & 3.327 & 4.357 & 4.742 & 3.044 & 2.513 & 2.231 & 2.039 & 2.234 & 2.20 & 2.10$^e$\tabularnewline
10 & 3.268 & 4.403 & 4.809 & 3.051 & 2.334 & 2.162 & 1.819 & 2.003 & 1.89 & \tabularnewline
12 & 3.293 & 4.414 & 4.865 & 3.036 & 2.289 & 2.150 & 1.699 & 1.890 &      & \tabularnewline
16 & 3.280 & 4.433 & 4.912 & 3.049 & 2.231 & 2.130 & 1.602 &       &      & \tabularnewline
20 & 3.276 & 4.439 & 4.926 & 3.052 &       & 2.125 & 1.576 &       &      & \tabularnewline
\hline
\end{tabular}

\begin{tablenotes}
 \item[a] Multireference perturbation theory calculations from \citeref{Nakayama1998_IJQC_157}.
 \item[b] Experimental value from \citerefs{Flicker1975_CPL_56} and \citenum{VanVeen1976_CPL_540}.
 \item[c] Experimental value from \citerefs{Mosher1973_CPL_332} and \citenum{Mosher1973_JCP_6502}.
 \item[d] Experimental value from \citerefs{Minnaard1973_RdTCdP_1179, Flicker1977_CPL_492, Kuppermann1979_CR_77}.
 \item[e] Experimental value from \citeref{Allan1984_HCA_1776}.
\end{tablenotes}
\end{threeparttable}%

\caption{Singlet-triplet gaps for polyenes in eV using RO-PPxr
  orbitals.\label{tab:stgaps-xpp}}
\end{table*}

\subsection{Ionization energies of polyenes}
Ionization energies in the polyenes are another interesting test of
the balance that various computational models can achieve for electron
correlation effects in doublet radical cations versus closed shell
neutrals. While modern DFT is generally acceptable for ionization
energies, DFT results for polyenes are considered unreliable in
general, because polyenes are known to exhibit strong static
correlation,\cite{Hachmann2006_JCP_144101, Hu2015_JCTC_3000,
  Lehtola2016_JCP_134110} and because DFT results for these systems
exhibit a strong dependence on the fraction of exact exchange, which
is often a symptom of delocalization error.\cite{Hait2018_JPCL_6280}

Pairing method calculations for the first vertical ionization energies
of the polyenes are shown in \cref{tab:polyene-ip-xpp}, again in the
quasi-full-valence active space. They are compared against
experimental values\cite{Plessis1987_CJP_165, Williams1991_JCP_6358,
  Ohno1995_JPC_14247, Mallard1983_JCP_5900, Bieri1977_HCA_2213,
  Allan1980_JCP_3114, Jones1979_IJMSIP_287} for the shorter polyenes,
as well as against domain localized pair natural orbital CCSD(T)
[DLPNO-CCSD(T)] calculations\cite{Bois2017_JCTC_4962} for the longer chain
species. Although the DLPNO-CCSD(T) calculations are extrapolated to
the complete basis set limit while the pairing active space
calculations exclude most dynamic correlations, it is nevertheless
striking that PQxr and PHxr agree roughly equally well with experiment
as the much more expensive CCSD(T) calculations. Unsurprisingly, the
simplest methods, PQ and PHQ, are in qualitative disagreement with the
higher-level calculations, and are less balanced than PP itself, as
they fail to correlate the unpaired electron. Even PQr is in error by
roughly 0.8 eV for the longer chain lengths, and it is striking how
much better-balanced the PQxr model is relative to the experiment and
full coupled-cluster results in these cases.

\begin{table*}
\begin{threeparttable}[t]
\setlength{\tabcolsep}{0.5em} 
\begin{tabular}{c|ccc|cc|ccc|c|c}
\toprule
$n_{C}$ & PP & PQ & PHQ & PQr  & PHrQ & PPxr & PQxr & PHxrQ   & CCSD(T)\tnote{a} & expt.\tabularnewline
\hline
2  & 9.528 & 10.078 & 10.154 & 10.056 & 10.127 & 9.508 & 10.049 & 10.126 &      & 10.51\tnote{b} \tabularnewline
4  & 8.490 &  9.374 &  9.562 &  9.103 &  8.981 & 8.213 &  8.791 &  8.903 & 9.27 & 9.08\tnote{c} \tabularnewline
6  & 7.883 &  8.822 &  9.089 &  8.379 &  8.206 & 7.418 &  7.930 &  8.091 &      & 8.29-8.30\tnote{d} \tabularnewline
8  & 7.569 &  8.603 &  8.870 &  8.132 &  7.827 & 7.081 &  7.536 &  7.676 & 7.93 & 7.79\tnote{e} \tabularnewline
12 & 7.182 &  8.310 &  8.581 &  7.814 &  7.364 & 6.676 &  7.051 &  7.159 & 7.32 &      \tabularnewline
16 & 6.998 &  8.178 &  8.431 &  7.681 &  7.138 & 6.491 &  6.832 &  6.893 & 6.96 &      \tabularnewline
20 & 6.905 &  8.118 &  8.348 &  7.621 &        & 6.398 &  6.728 &        & 6.74 &      \tabularnewline
24 & 6.857 &  8.090 &        &  7.593 &        & 6.350 &  6.676 &        & 6.57 &      \tabularnewline
\hline
\end{tabular}

\begin{tablenotes}
 \item[a] DLPNO-CCSD(T) calculations extrapolated to the basis set
   limit from \citeref{Bois2017_JCTC_4962}.
 \item[b] Experimental value from \citerefs{Plessis1987_CJP_165,
   Williams1991_JCP_6358, Ohno1995_JPC_14247}.
 \item[c] Experimental value from \citeref{Mallard1983_JCP_5900}.
 \item[d] Experimental value from \citerefs{Bieri1977_HCA_2213} and \citenum{Allan1980_JCP_3114}.
 \item[e] Experimental value from \citeref{Jones1979_IJMSIP_287}.

\end{tablenotes}
\end{threeparttable}%

\caption{First ionization energies of polyenes in eV using RO-PPxr
  orbitals. \label{tab:polyene-ip-xpp}}
\end{table*}

\subsection{Doublet-quartet gaps of polyenyls}
The calculated spin gap between the doublet (ground) state and the
quartet (excited) state of the polyenyl series is shown in
\cref{tab:dqgaps-xpp}, as a function of the number of C atoms in the
radical chain. These gaps do not appear to have been thoroughly
investigated: as far as we are aware, doublet-quartet gaps are only
available for \ce{C3H5} from CASPT2 calculations
(\citeref{Aquilante2003_CPL_689}) and from experiment (\citeref{Fischer2002_JPCA_4291}),
and so we only report data for the three smallest polyenyls.  Although
there is a slight discrepancy from experiment, likely arising from the
already mentioned issues of basis set and missing dynamical
correlation, the radical models are, however, in good agreement with
the CASPT2 value of \citeref{Aquilante2003_CPL_689} for \ce{C3H5}. The
original PP, PQ, and PHQ models not only show a larger error for
\ce{C3H5}, but also do not predict monotonic behavior for the
doublet-quartet gap as they do not correlate the radical electrons at
all.

Calculations for the first vertical ionization energies of the
polyenyls are shown in \cref{tab:polyenyl-ip-xpp}, again in the
quasi-full-valence active space. They are compared against
experimental values from \citerefs{Griller1981_JACS_1586,
  Houle1978_JACS_3290, Lossing1976_IJMSIP_9,
  Pignataro1967_JACS_3693}. The radical models are once again in good
agreement with experiment, showing monotonically decreasing ionization
energies with increasing carbon chain length.

\begin{table*}
\begin{threeparttable}[t]
\setlength{\tabcolsep}{0.5em} 
\begin{tabular}{c|ccc|cc|ccc|c|c}
\toprule
$n_{C}$ & PP & PQ & PHQ & PQr  & PHrQ & PPxr & PQxr & PHxrQ   & CASPT2\tnote{a} & expt.\tnote{b}\tabularnewline
\hline
3  & 4.691 & 5.131 & 5.190 & 5.330 & 5.562 & 4.905 & 5.460 & 5.585 & 5.89 & 6.33 \tabularnewline
5  & 4.779 & 5.377 & 5.497 & 5.319 & 5.083 & 4.755 & 4.930 & 4.978 &  & \tabularnewline
7  & 2.917 & 3.705 & 3.960 & 3.174 & 3.227 & 2.469 & 2.926 & 3.140 &  & \tabularnewline
\hline
\end{tabular}

\begin{tablenotes}
 \item[a] Calculation from \citeref{Aquilante2003_CPL_689}.
 \item[b] Experimental value from \citeref{Fischer2002_JPCA_4291}.
\end{tablenotes}
\end{threeparttable}%

\caption{Doublet-quartet gaps for polyenyls in eV using RO-PPxr
  optimized orbitals.\label{tab:dqgaps-xpp}}
\end{table*}

\begin{table*}
\begin{threeparttable}[t]
\setlength{\tabcolsep}{0.5em} 
\begin{tabular}{c|ccc|cc|ccc|c}
\toprule
$n_{C}$ & PP & PQ & PHQ & PQr  & PHrQ & PPxr & PQxr & PHxrQ & expt.\tabularnewline
\hline
3 & 6.996 & 7.088 & 7.058 & 7.506 & 7.742 & 7.403 & 7.773 & 7.804 & 8.1\tnote{a}, 8.13\tnote{b}\tabularnewline
5 & 6.043 & 6.065 & 5.995 & 6.692 & 6.999 & 6.671 & 7.081 & 7.094 & 7.25\tnote{c}, 7.76\tnote{d} \tabularnewline
7 & 5.544 & 5.763 & 5.610 & 6.374 & 6.733 & 6.119 & 6.894 & 6.869 & \tabularnewline
9 & 5.290 & 5.307 & 5.145 & 6.011 & 6.412 & 5.967 & 6.592 & 6.558 & \tabularnewline
\hline
\end{tabular}

\begin{tablenotes}
\item[a] Experimental value from \citeref{Griller1981_JACS_1586}.
\item[b] Experimental value from \citeref{Houle1978_JACS_3290}.
\item[c] Experimental value from \citeref{Lossing1976_IJMSIP_9}.
\item[d] Experimental value from \citeref{Pignataro1967_JACS_3693}.
\end{tablenotes}
\end{threeparttable}%

\caption{First ionization energies of polyenyls in eV, computed using
  PPxr orbitals.\label{tab:polyenyl-ip-xpp}}
\end{table*}

\subsection{Optimal orbitals}
Finally, in addition to the study of the numerical performance of the
models, it is also interesting to examine the resulting orbital
picture. The neutral polyenes and charged polyenyl species have
singlet ground states in which all electrons are paired. Here, we will
examine the orbitals for the unpaired electron in a charged polyene,
\ce{C20H22+}, and a neutral polyenyl radical, \ce{C21H23}.

\begin{figure*}
  \subfloat[Molecular geometry]{\includegraphics[width=1\textwidth]{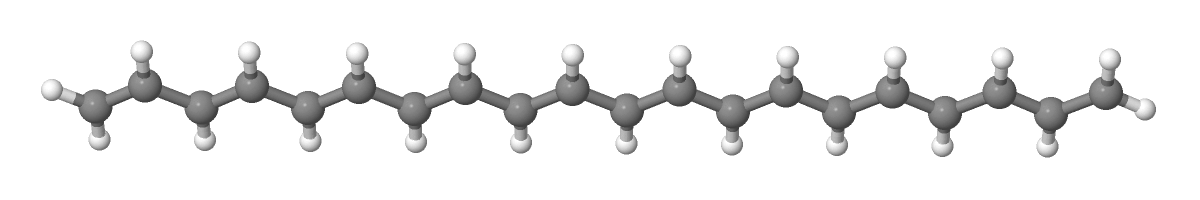}}

  \subfloat[ROHF]{\includegraphics[width=1\textwidth]{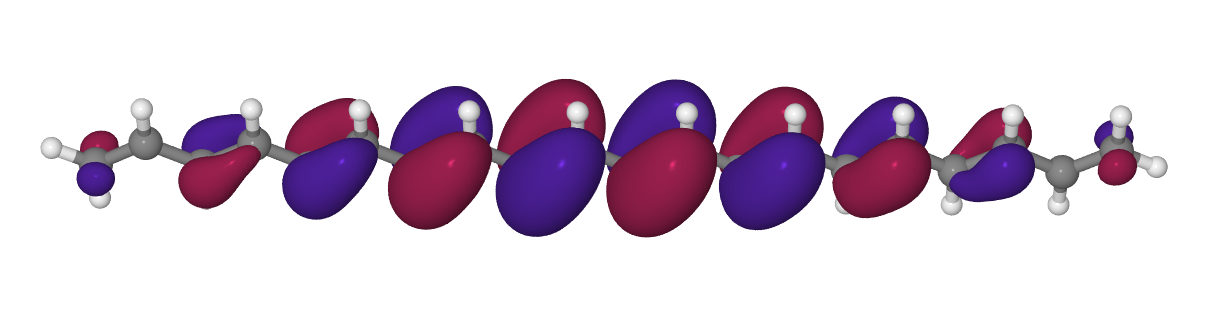}}

  \subfloat[PP]{\includegraphics[width=1\textwidth]{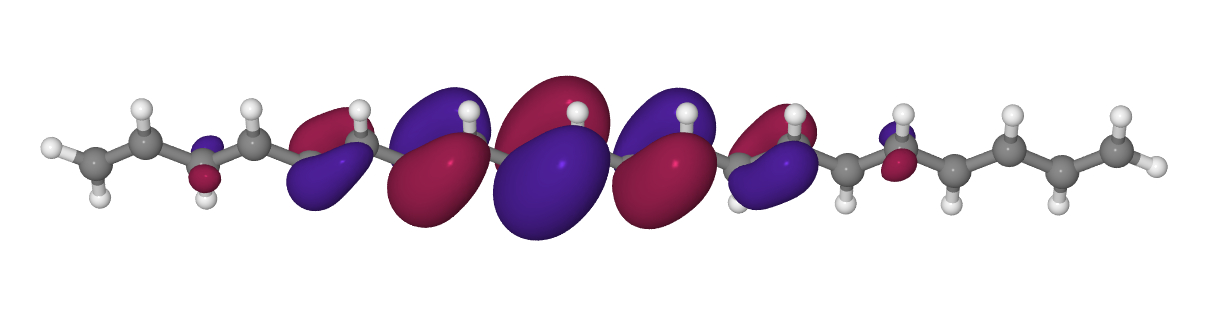}}

  \subfloat[PPxr]{\includegraphics[width=1\textwidth]{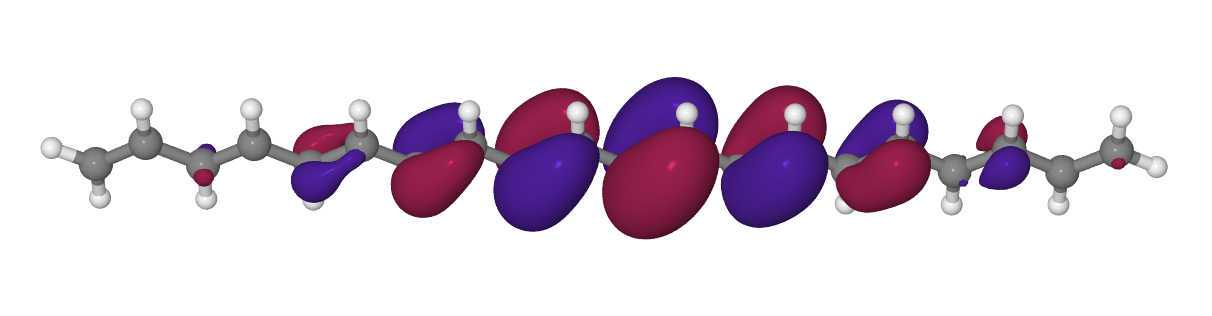}}

  \caption{The 85 \% density containment plot \cite{Haranczyk2008_JCTC_689,
      Lehtola2014_JCTC_642} of the unpaired, singly occupied orbital
    in \ce{C20H22+} in the ROHF, PP, and PPxr
    models. A ROHF calculation with Q-Chem\cite{Epifanovsky2021_JCP_84801} converges to the same energy, and stability analysis showed the ROHF solution to be a proper local minimum. \label{fig:orbital-cation}}
\end{figure*}

The unpaired $\pi_u$ orbital in \ce{C20H22+} corresponding to the
ROHF, PP, and PPxr levels of theory are shown in
\cref{fig:orbital-cation} using an 85\% density containment criterion to define the isosurfaces.
\cite{Haranczyk2008_JCTC_689, Lehtola2014_JCTC_642} The ROHF SOMO is delocalized over
the whole chain and has inversion symmetry (a$_\mathrm{u}$ irreducible representation) about the center of the
molecule, which is consistent with the C$_\mathrm{2h}$ nuclear framework symmetry. Additional ROHF calculations followed by stability analysis using Q-Chem\cite{Epifanovsky2021_JCP_84801} confirm that this solution is a local minimum. Surprisingly, even though the unpaired orbital is not
correlated at the PP level of theory, PP partially localizes the SOMO around
the short C(9)--C(10) bond (numbering the atoms from the left), to
cover just over half the chain, while breaking point group symmetry. The PPxr SOMO is relatively little changed from PP. The main visual difference is that it localizes on the short C(11)--C(12) bond. However, that is symmetry
equivalent to the C(9)--C(10) bond about which the PP orbital localized.  
These findings on the degree of 
localization of the unpaired singly occupied orbital are in good
qualitative agreement with a previous study employing the
orbital-optimized scaled opposite-spin MP2 method.\cite{Kurlancheek2012_JCP_54113}

\begin{figure*}
  \subfloat[Molecular geometry]{\includegraphics[width=1\textwidth]{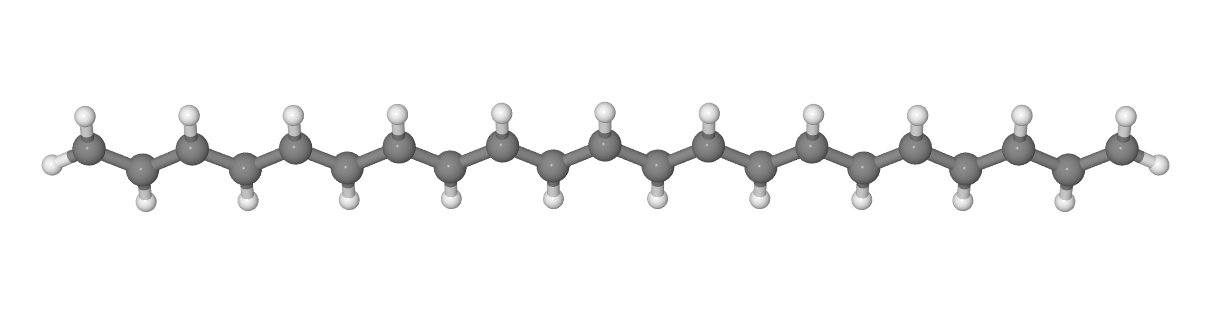}}

  \subfloat[ROHF]{\includegraphics[width=1\textwidth]{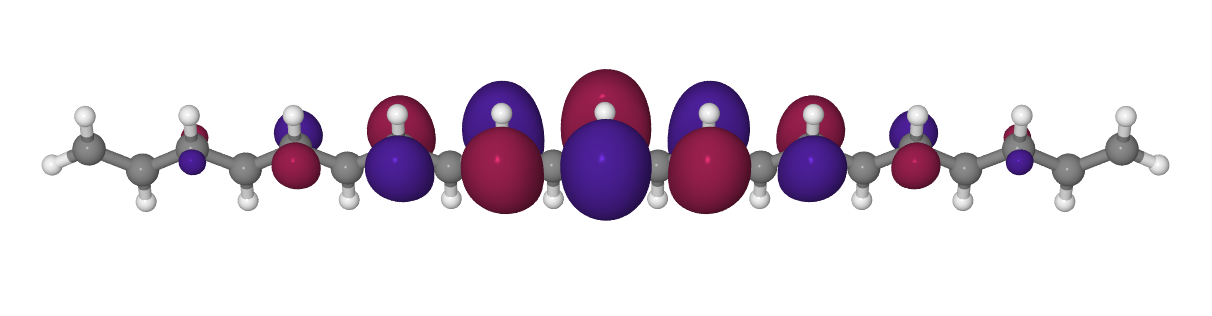}}

  \subfloat[PP]{\includegraphics[width=1\textwidth]{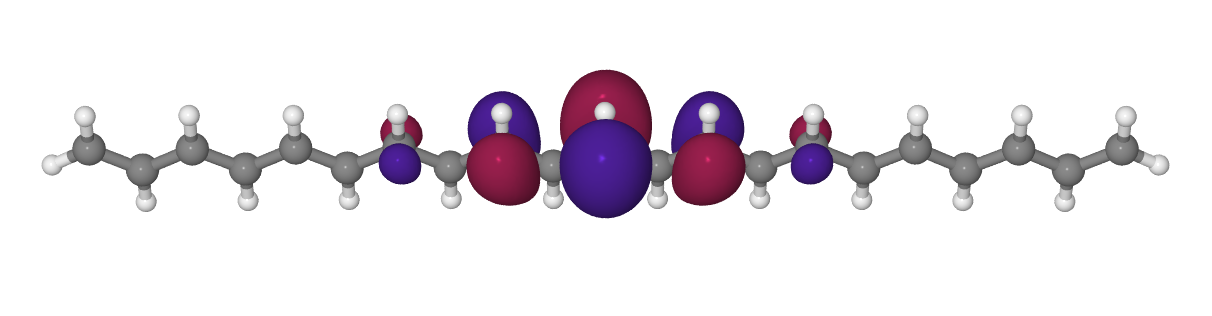}}

  \subfloat[PPxr]{\includegraphics[width=1\textwidth]{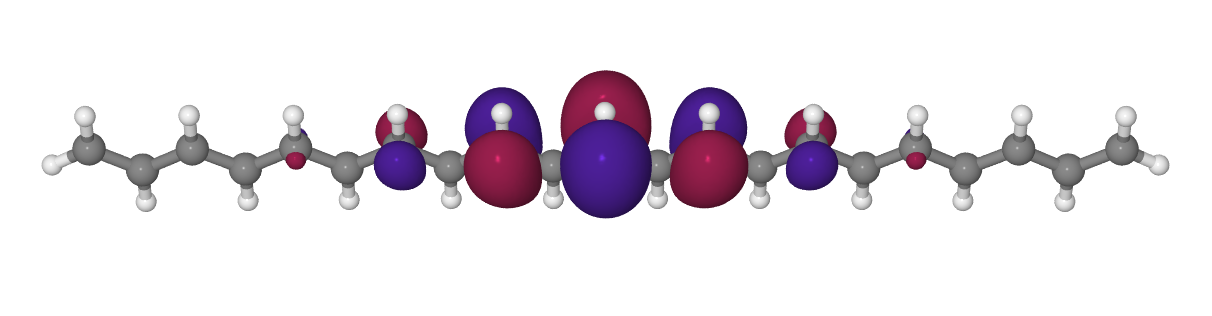}}

  \caption{The 85 \% density containment plot \cite{Haranczyk2008_JCTC_689,
      Lehtola2014_JCTC_642} of the unpaired, singly occupied orbital in
    \ce{C21H23} in the ROHF, PP, and PPxr
    models. A ROHF calculation with Q-Chem converges to the same energy, and stability analysis showed the ROHF solution to be a proper local minimum.\label{fig:orbital-neutral}}
\end{figure*}

Repeating the demonstration for the unpaired SOMO in
\ce{C21H23} (which has C$_\mathrm{2v}$ framework symmetry; \cref{fig:molstruct}), one obtains the results shown in
\cref{fig:orbital-neutral}. The ROHF orbital is again delocalized over
almost the whole molecule, though to a lesser extent than in the polyene
cations. Symmetry is again preserved. In PP and PPxr, the radical electron localizes strongly
around the two single bonds at the center of the molecule, the PPxr
orbital being slightly less localized as expected due to the
correlations with the paired electrons in the model. In contrast to the polyene cation case, the PP and PPxr SOMOs preserve point group symmetry. This reflects our choice of model geometry, which was designed for radicals in the case of the polyenyl chains, but for closed shell neutrals in the polyene chains. Overall both of these examples illustrate the fact that correlation effects induce partial localization of the radical electron relative to mean-field ROHF.

\section{Summary and Discussion}
\label{sec:summary}

We have presented the extensions of the perfect pairing (PP), the perfect quadruples (PQ) and the CCSDTQ subset of the the perfect hextuples (PH) models to open-shell
systems, and demonstrated their accuracy with calculations on the
ground, excited and cationic states of polyenes and polyenyls. The results are
encouraging, as they indicate the feasibility of accurate yet
cost-efficient \emph{ab initio} models for open-shell systems
exhibiting strong correlation.

Although the pairing models are limited to symmetric active spaces of
$N$ electrons in $N$ orbitals and not all strong correlation problems
are amenable to a single-reference coupled-cluster description upon
which the pairing models rely,\cite{Lehtola2018_MP_547, Lehtola2017_JCP_154105} the pairing models
are ideal for the description of large hydrocarbons where the natural
full-valence active space is $N$ electrons in $N$ orbitals and the strong correlation effects appear to be describable by a truncated CCSDTQ or CCSDTQ56 model as found in this work and \citerefs{Lehtola2016_JCP_134110} and \citenum{Lehtola2018_MP_547}.

The approach outlined in the present work could also be easily used to
extend the pairing models to e.g. asymmetric active spaces by
introducing new classes depicting lone electron pairs of electron-rich
atoms or vacant orbitals in electron-poor atoms, or further to
dynamical correlation by introducing further excitations to the
inactive virtual orbitals.  While such extensions might be very
appealing for applications on chemical problems, the challenge arises
in that the favorable scaling of the pairing approaches is destroyed
by the additional external labels whose size increase with system
size, unlike the present case.  In order to tackle these cases, one
alternative might be to omit the pairing altogether but restrict the
realm of the higher connected excitation operators as in active-space
coupled-cluster methods.\cite{Piecuch1999_JCP_6103, Olsen2000_JCP_7140}

As we have numerically demonstrated here and in our previous
work,\cite{Lehtola2018_MP_547} orbital-optimized coupled-cluster theory does
not converge to the full configuration interaction (FCI) limit in the
absence of single excitations.\cite{Koehn2005_JCP_84116} However, the pairing
models are straightforward to extend to a non-orthogonal
coupled-cluster treatment which does converge to the FCI
limit.\cite{Myhre2018_JCP_94110} Non-orthogonal
extensions of the pairing models could be investigated in future work.

\clearpage

\section*{Acknowledgments}

\edit{We take the opportunity to
  acknowledge the many friendly and very valuable scientific discussions
  with Piotr Piecuch about numerous aspects of quantum chemistry,
  particularly coupled cluster theory, over the past three decades. We look forward to many more!}  This work was supported by the
Director, Office of Basic Energy Sciences, Chemical Sciences,
Geosciences, and Biosciences Division of the U.S.  Department of
Energy, under Contract No. DE-AC02-05CH11231, as well as the Academy
of Finland under grant numbers 311149, 350282, and
353749. Computational resources provided by CSC -- It Center for
Science Ltd (Espoo, Finland) and the Finnish Grid and Cloud
Infrastructure (persistent identifier
urn:nbn:fi:research-infras-2016072533) are gratefully acknowledged.
\bibliography{citations,mhgxtras}

\end{document}